\documentclass[10pt,twocolumn]{article}

\usepackage[utf8]{inputenc}
\usepackage[T1]{fontenc}
\usepackage{times}
\usepackage[margin=0.75in,top=1in,bottom=1in]{geometry}
\usepackage{graphicx}
\usepackage{grffile}
\usepackage{booktabs}
\usepackage{array}
\usepackage{tabularx}
\usepackage{multirow}
\usepackage{caption}
\usepackage{subcaption}
\usepackage{amsmath}
\usepackage{hyperref}
\usepackage{xcolor}
\usepackage{url}
\makeatletter
\g@addto@macro{\UrlBreaks}{\UrlOrds}
\makeatother
\usepackage{titlesec}
\usepackage{enumitem}
\usepackage{fancyhdr}
\usepackage{float}
\usepackage{makecell}
\usepackage{textcomp}
\usepackage{authblk}
\usepackage{abstract}
\usepackage{adjustbox}
\usepackage{dblfloatfix}
\usepackage{ragged2e}
\RaggedRight
\usepackage{placeins}

\hypersetup{colorlinks=true,linkcolor=blue!60!black,citecolor=blue!60!black,urlcolor=blue!60!black}
\titleformat{\section}{\large\bfseries}{\thesection.}{0.5em}{}
\titleformat{\subsection}{\normalsize\bfseries}{\thesubsection}{0.5em}{}
\titlespacing*{\section}{0pt}{1.2ex plus 0.5ex minus .2ex}{0.8ex plus .2ex}
\titlespacing*{\subsection}{0pt}{1.0ex plus 0.4ex minus .2ex}{0.6ex plus .2ex}
\newcommand{\figframe}[1]{\setlength{\fboxsep}{0pt}\setlength{\fboxrule}{0.4pt}\fbox{#1}}

\begin{document}

\twocolumn[
\begin{@twocolumnfalse}
\begin{center}
{\LARGE\bfseries Reduced-Mass Orbital AI Inference via Integrated\\[4pt] Solar, Compute, and Radiator Panels\par}
\vspace{10pt}
{\large Stephen Gaalema, Samuel Indyk, and Clinton Staley\\[3pt]
\textit{University of Austin}}
\vspace{6pt}

{\small\textit{Acknowledgements: Dr.\ Michael Moffitt, Tyler Erickson}}
\end{center}
\vspace{6pt}
\begin{onecolabstract}
We describe and analyze a distributed compute architecture for SSO computational satellites that can potentially provide $>$100~kW compute power per launched metric ton (including deployment and station keeping mass). The architecture colocates and integrates the solar cells, radiator, and compute functions into multiple small panels arranged in a large array. The resultant large vapor chamber radiator area per panel should permit ICs to operate at junction temperatures near 40\textdegree C with benefits in compute efficiency and reliability. Using the structure of the radiator to support the solar cells may also yield a specific power of about 500~W/kg compared to less than 100 for existing conventional implementations.

Assuming development of custom solutions for all components, a full 16 MW computational satellite (150 tons) comprising a 20\,m~$\times$~2200\,m grid of 16,000 panels can fit in a single Starship hold
The concept is scalable to much larger satellites with higher mass payloads or using on-orbit assembly. We consider panel sizes from 1 to 4~m\textsuperscript{2} to allow trading vapor chamber heat transport with compute efficiency and inter-panel communication.

Assuming a 1\,kW/panel design, 512-panel subarrays of the satellite can run a representative inference-only LLM with 500,000 token context window and 128 attention blocks, at a rate of 553 tokens/sec/session, across 256 simultaneous in-flight sessions. A full satellite could support 31 such subarrays, for $>$7900 inferences at a time.
\end{onecolabstract}
\vspace{10pt}
\end{@twocolumnfalse}
]

\section{Introduction}

Orbital data centers are being proposed and even tested at small scale. They offer the promise of scalability not available on Earth, due to almost unlimited access to continuous solar power in a dawn-dusk SSO, and avoid issues with terrestrial data centers such as demands on power and water resources, opposition from environmental groups, and permitting delays.

If cost and other issues are resolved, construction of these data centers could dominate the rapidly expanding market for AI computation, potentially trillions of dollars per year. With the immense resources likely to be expended, development of highly optimized system designs and cost critical custom components (solar cell arrays, compute chips and packaging, heat radiators) for the satellites is expected.

Current work generally favors system partition similar to existing satellites with separate structures for solar power, compute, and thermal radiator:

\begin{itemize}[nosep,leftmargin=1.5em]
    \item large solar cell arrays (possibly even kilometer scale) providing power
    \item a centralized compute structure of minimum size to reduce compute latency and interconnect power
    \item a two sided heat radiator operating at a high temperature to minimum size and reduce mass
\end{itemize}

We consider a very different system concept that may allow lower cost space data centers than would be expected from extrapolating current technology.

We propose compute and radiator functions distributed over the same area as the solar arrays to reduce mass and radiator complexity. We limit initial scope to satellites in dawn-dusk sun synchronous orbit (SSO) with mass and stowed volume compatible with single launch capabilities of the SpaceX V3 or V4 Starship expected in the next few years.

We assume:

\begin{itemize}[nosep,leftmargin=1.5em]
    \item solar cells with 20\% to 27\% efficiency providing up to 370~W/m\textsuperscript{2} power
    \item compute modules distributed to match the solar cell power output density
    \item a single sided heat radiator ideally operating at 20 to 30\textdegree C, feasible at 370~W/m\textsuperscript{2}.
    \item compute implemented with ICs optimized for operation at junction temperatures roughly 5\textdegree C less than the back radiator temperatures.
\end{itemize}

The two key advantages of this ``Integrated Solar Compute Radiator'' (ISCR) architecture over high temperature radiator designs are:

\begin{itemize}[nosep,leftmargin=1.5em]
    \item \textbf{Compute speed and power efficiency:} low junction temperatures allow somewhat higher clock rates, lower supply voltages, and much lower leakage currents.
    \item \textbf{Structural:} each 1--2\,m square panel integrates solar cells on its front face, an insulating gap, and a vapor chamber backplane that simultaneously serves as the sole mechanical substrate for the solar cells and as the thermal radiator. Compute ICs are sandwiched between the solar layer and the radiator --- no dedicated solar panel structure and associated power distribution is required.
\end{itemize}

The ISCR satellite comprises thousands of such panels arranged in a large linear array, providing distributed compute that addresses thermal management without moving parts.

Table~\ref{tab:temps} shows front and back face temperatures for several orbital positions of interest. In dawn-dusk sun-synchronous orbit (SSO), the satellite's orbital plane remains perpendicular to the Earth-Sun line throughout the year, providing continuous solar illumination without eclipse. Earth's infrared emission adds a modest back-face heat load that increases at lower altitudes, but even at 600\,km SSO the back face remains below 24\textdegree C under near-ideal assumptions.

\begin{table}[!ht]
\centering
\caption{Temperatures for front and back radiators with nearly ideal assumptions (emissivity 0.92, insulated from each other, triple junction GaAs solar cells) for various possible orbits affected by IR radiation from Earth).}
\label{tab:temps}
\small
\begin{adjustbox}{max width=\columnwidth}
\begin{tabular}{lcccc}
\toprule
 & \textbf{Deep Space} & \textbf{SSO} & \textbf{SSO} & \textbf{SSO} \\
 & \textbf{(L5)} & \textbf{(2000\,km)} & \textbf{(1000\,km)} & \textbf{(600\,km)} \\
\midrule
\textbf{Front Temp} & 84.3\textdegree C & 85.7\textdegree C & 86.7\textdegree C & 87.7\textdegree C \\
\textbf{Back Power Load} & 368\,W & 382\,W & 392\,W & 402\,W \\
\textbf{Back Temp} & 17.2\textdegree C & 20.0\textdegree C & 22.0\textdegree C & 23.6\textdegree C \\
\bottomrule
\end{tabular}
\end{adjustbox}
\end{table}

The baseline design also assumes 1000\,km SSO, where the orbital plane is maintained perpendicular to the Earth--Sun line throughout the year, yielding continuous solar illumination without eclipse periods. But higher orbits or deep space may be of interest if environmental issues limit use of LEO.

For this analysis, we consider the integrated solar/compute/radiator (ISCR) array divided into panels with sizes from 1 to 4~m\textsuperscript{2}. Large sizes are constrained by the radiator panel vapor chamber efficiency. Most calculations will assume a panel size that provides 1\,kW of power to the compute module.

Design of a system as complex as a 16\,MW orbital data center satellite while allowing almost all components to be custom optimized for the specific application is an opportunity of a scale that rarely exists, and is a huge undertaking. Consequently, our conceptual design presented here may use best-estimate assumptions and estimates that we intend to refine with future work and outside critique.

\section{Related Work}

The following section summarizes key related works in space-based AI compute infrastructure. Each has informed aspects of the ISCR design.

\textbf{Google Project Suncatcher.} Google's research preprint~\cite{suncatcher} proposes ``Project Suncatcher'' --- compact constellations of solar-powered satellites carrying Google TPUs connected by free-space optical inter-satellite links in a dawn-dusk sun-synchronous orbit. Their work addresses three core challenges: (a) achieving data-center-scale inter-satellite bandwidth via DWDM optical links (demonstrated 1.6~Tbps per transceiver pair on a bench-scale demonstrator); (b) controlling tightly-clustered satellite formations at km-scale separations; and (c) radiation hardening --- they conducted radiation tests on TPUs targeting a 5-year LEO operational lifetime. Their proposed 81-satellite formation at 1-km radius represents a constellation approach, contrasting with our single-satellite linear panel design. Key takeaway: Google confirms $>$1~Tbps optical inter-satellite links are feasible at km-scale separations --- our design achieves comparable bandwidth (800~Gb/s) at 0.5\,m via copper, solving the link budget problem differently.

\textbf{Bargatin et al.\ --- Tether-Based Orbital AI Data Centers.} The University of Pennsylvania group~\cite{bargatin} proposes a vertical tether architecture spanning tens of kilometers in dawn-dusk SSO. Their two reference designs ($\sim$2~MW and $\sim$20~MW) achieve continuous solar illumination by aligning with gravity-gradient forces along the tether. Each compute node includes integrated radiators, with a passive solar-pressure attitude control scheme. Their work is the most comprehensive system-level design published for tether-based orbital data centers and provides excellent validation of the physics of large-structure orbital compute. Our design operates at a similar scale (16~kW vs 2--20~MW), but targets a single-launch per satellite and near-term deployable configuration.

\textbf{Sophia Space.} Sophia Space~\cite{sophia} is a commercial startup developing flat ``TILE'' modules that integrate solar cells, compute, and passive radiative cooling. Their SOOS (Sophia Orbital Operating System) manages a modular array of up to 2,500 tiles. Key figures: 92\% compute power utilization (vs $\sim$50--60\% terrestrial after cooling overhead), 30-year service life, passive radiative cooling to deep space eliminating water/HVAC. Their patent (US~20250108938) details tile arrays with integrated heat dissipation, compute modules, and scalability to 600~Exa-ops at 100~MW. Sophia Space's per-tile approach directly validates our per-panel integrated design philosophy, but does not deeply analyze implementation of an LLM or other deep learning architecture on an array of such panels.

\textbf{Vicinanza --- Industry Survey.} An industry survey~\cite{vicinanza} reviews Google's Suncatcher, Musk/Starlink space compute hints, and key challenges: thermal (radiator mass), reliability (solar degradation, radiation), and latency ($>$100~ms to ground). The survey argues that AI training is best suited to space compute (continuous solar) while inference may stay terrestrial --- we argue the opposite: inference (latency-tolerant for asynchronous queries) is the near-term space compute opportunity. Training requires tight synchronization incompatible with inter-panel latency.

\section{Background}

This section provides relevant background on solar cells, vapor chamber radiators, and LLM parallelism, which we draw on in the detailed design sections that follow.

\subsection{Solar Cells}

Solar cell technology relevant to the ISCR architecture spans a range of efficiency-mass-cost tradeoffs. Three technologies are of primary interest: perovskite/thin-Si tandem, triple junction GaAs, and space-grade crystalline silicon. Table~\ref{tab:solar} compares representative designs at high temperatures we expect, and includes the ISS iROSA panel and an estimated Starlink~V3 configuration at typical temperatures for context. Cost numbers are extremely variable and only provided as very rough estimates.

\begin{table*}[!ht]
\centering
\caption{Comparison of solar cell and panel designs.}
\label{tab:solar}
\small
\begin{tabular}{lcccc}
\toprule
\textbf{Solar panels} & \textbf{Cost-performance} & \textbf{High} & \textbf{ISS iROSA} & \textbf{estimated Starlink V3} \\
 & \textbf{balance} & \textbf{performance} & \textbf{(Single Panel)} & \textbf{(Single-Satellite)} \\
\midrule
cell type & Perovskite/50\,$\mu$m Si & Triple-junction GaAs & Triple-junction GaAs & Space-grade thick Si \\
present cost, \$/W & --- & 20--80 & 50--100 & 10--30 \\
future cost, \$/W & 5--15 & 10--30 & 20--60 & 5--20 \\
solar cell size, mm & flexible & flexible & 62$\times$127 & --- \\
cell temperature, \textdegree C & 70 & 87 & 32 & 32 \\
Cell Efficiency & 27\% & 27\% & 30.7\% & 18\% \\
W/m\textsuperscript{2} in SSO & 363 & 369 & 407 & 239 \\
Total area density, kg/m\textsuperscript{2} & \textbf{0.5} & \textbf{0.5} & 6 & --- \\
Specific power, W/kg & 506 & 506 & 61.5 & 50 \\
Power per ton, kW/ton & \textbf{726} & \textbf{738} & 40.6 & --- \\
Power Output, kW & \textbf{10890} & \textbf{11073} & 28 & 59.7 \\
Deployment & Flexible roll-out & Flexible roll-out & Flexible roll-out & Rigid fold-out \\
Stowed power density, kW/m\textsuperscript{3} & 22.2 & 22.6 & 40 & 15 \\
Stowed density, tons/m\textsuperscript{3} & 0.03 & 0.03 & --- & --- \\
\bottomrule
\end{tabular}
\end{table*}

Cost and mass for the ISS proven GaAs is likely prohibitive, although thin GaAs (ELO) would address mass, even with development, cost is likely still an issue. The lower cost alternatives are considered below.

\begin{figure}[!ht]
\centering
\figframe{\includegraphics[width=\linewidth]{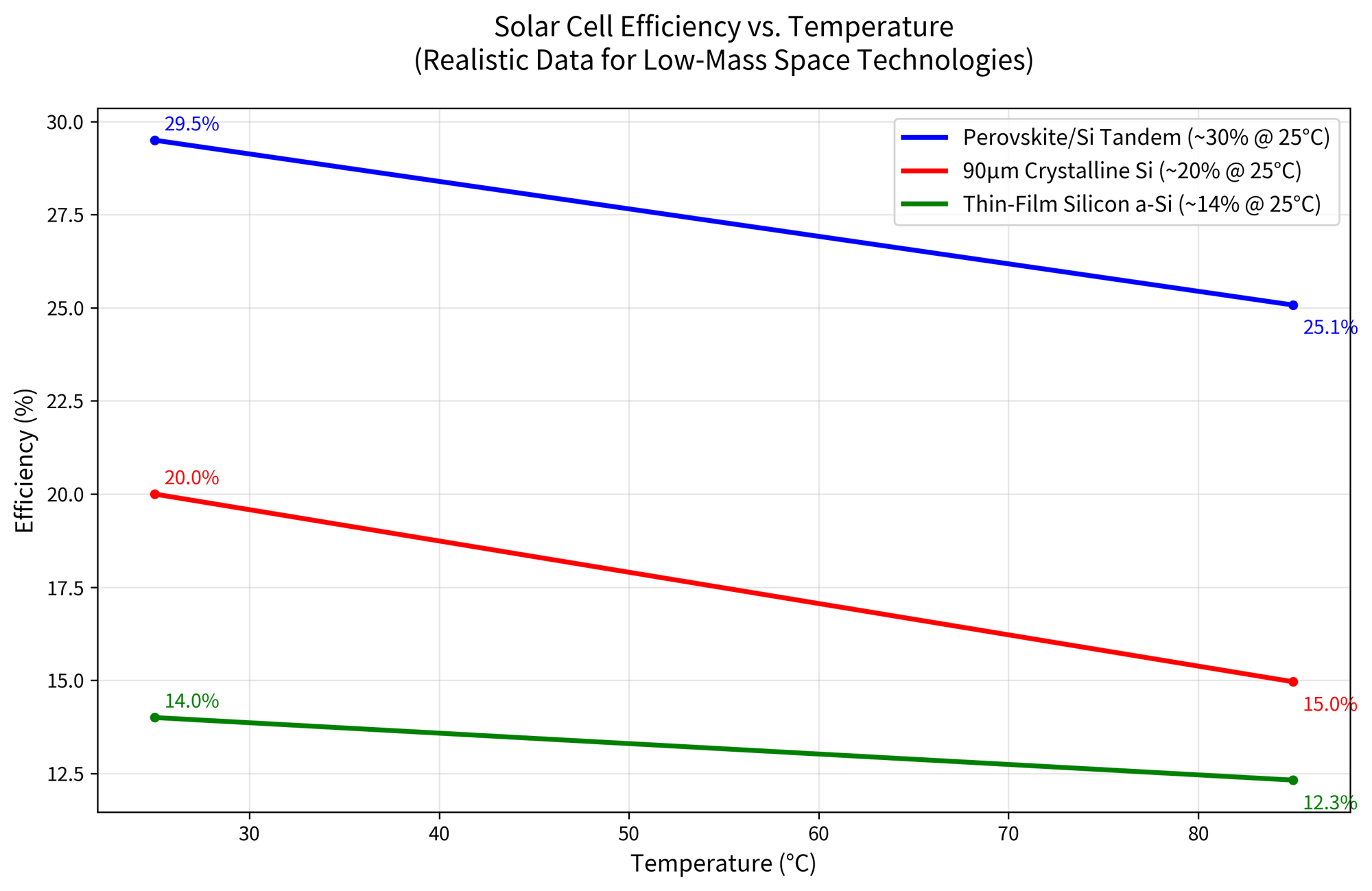}}
\caption{Efficiency vs.\ Temperature for three low mass solar cell technologies across the 40--85\textdegree C range typical of orbital operation. Perovskite/Si Tandem ($\sim$30\% @ 25\textdegree C) maintains the highest efficiency ($\sim$25.4\% at 85\textdegree C)~\cite{chen2024,oxfordpv}. 90\,$\mu$m Crystalline Si ($\sim$20\% @ 25\textdegree C) declines to $\sim$15.5\% at 85\textdegree C~\cite{yoshikawa2023,nrel2024}. Thin-Film Silicon a-Si ($\sim$14\% @ 25\textdegree C) is the most thermally stable but lowest-efficiency option, ending at $\sim$11.9\% at 85\textdegree C~\cite{sai2024,iec2024}.}
\label{fig:solar-efficiency}
\end{figure}

For missions requiring high power density in hot environments, the Perovskite/Si Tandem is the clear choice among these.

\textbf{Why the mass is so low:} The Perovskite layer is a direct bandgap material with a very high absorption coefficient, requiring only $\sim$400\,nm thickness to capture high-energy photons --- less than 2~g/m\textsuperscript{2} added weight. Combined with a 50~$\mu$m silicon layer, the tandem achieves specific powers exceeding 1.0~kW/kg. For a 2\,m~$\times$~2\,m panel, the mass breakdown is approximately:

\begin{itemize}[nosep,leftmargin=1.5em]
    \item Perovskite + Silicon (50\,$\mu$m): $\sim$0.12~kg/m\textsuperscript{2}
    \item Polyimide Substrate (Kapton): $\sim$0.05~kg/m\textsuperscript{2}
    \item Interconnects + Encapsulation: $\sim$0.15~kg/m\textsuperscript{2}
    \item Total Mass: $\sim$0.32~kg/m\textsuperscript{2} without a protective layer
\end{itemize}

\textbf{Support substrate.} The most common space-grade flexible substrate is Kapton (Polyimide), available at 25--50~$\mu$m thickness (adding only 35--70~g/m\textsuperscript{2}), stable from $-$269\textdegree C to $+$400\textdegree C. \textbf{Structural reinforcement} uses carbon fiber mesh or CFRP skins to prevent 50~$\mu$m silicon from cracking during maneuvers while remaining lightweight. \textbf{Interconnects} use flexible flat cables (FFC) for low-torque deployment and interdigitated foil interconnects for rear-contact cells, eliminating front-face shadowing.

\begin{table}[!ht]
\centering
\caption{Component materials for 50\,$\mu$m thick Solar Cells.}
\label{tab:materials}
\small
\begin{adjustbox}{max width=\columnwidth}
\begin{tabular}{lll}
\toprule
\textbf{Component} & \textbf{Recommended Material} & \textbf{Benefit for 50\,$\mu$m Cells} \\
\midrule
Main Substrate & Polyimide (Kapton) & \makecell[l]{High dielectric strength;\\handles $-$269\textdegree C to $+$400\textdegree C.} \\
\addlinespace
Structural Backbone & Carbon Fiber Mesh & \makecell[l]{High stiffness; prevents\\micro-cracking in silicon.} \\
\addlinespace
Electrical Path & Metal Foil / FFCs & \makecell[l]{Ultra-low profile; allows\\for 5\,cm bend radii.} \\
\addlinespace
Protective Layer & ETFE / Ceramic Film & \makecell[l]{Replaces heavy glass;\\protects against Atomic\\Oxygen (ATOX).} \\
\bottomrule
\end{tabular}
\end{adjustbox}
\end{table}

\subsection{Vapor Chambers}

A vapor chamber is a sealed, flat heat spreader containing a working fluid and a wick structure. Heat applied at a localized source --- in the ISCR, the compute package --- evaporates the working fluid, which migrates as vapor to cooler regions and condenses, with the liquid returned to the heat source via capillary action in the wick. This two-phase cycle can spread heat over distances up to approximately one meter, enabling near-uniform temperature across the full panel back face despite a single, small heat source.

\begin{figure*}[!ht]
\centering
\begin{subfigure}[t]{0.48\textwidth}
    \centering
    \figframe{\includegraphics[width=\linewidth]{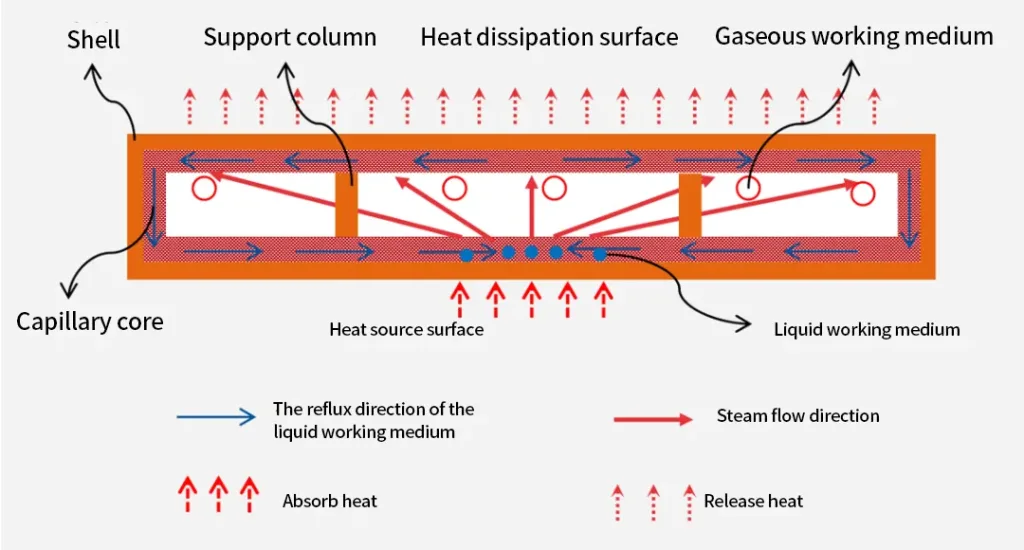}}
    \caption{}
\end{subfigure}
\hfill
\begin{subfigure}[t]{0.48\textwidth}
    \centering
    \figframe{\includegraphics[width=\linewidth]{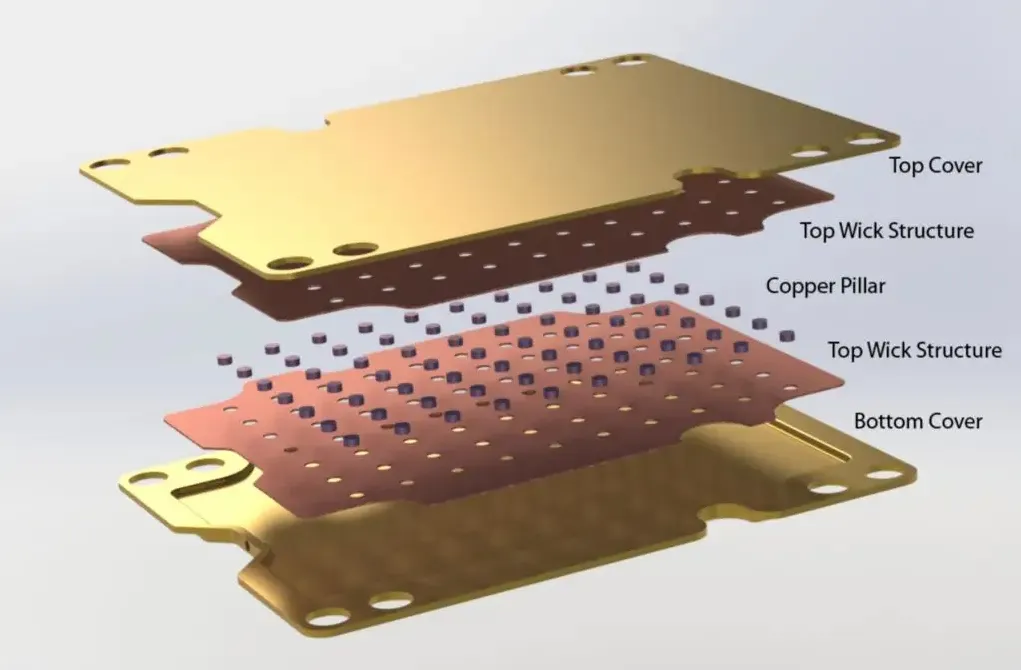}}
    \caption{}
\end{subfigure}
\caption{(a)~Heat flow in a vapor chamber showing evaporation at the heat source, vapor transport, condensation at cooler regions, and capillary return of liquid. (b)~Exploded view of a vapor chamber assembly, showing the layered construction: top cover, top wick (sintered/mesh copper), copper pillar array providing structural support and vapor channels, bottom wick, and bottom cover. The pillar array maintains chamber thickness while allowing vapor and liquid flow.~\cite{ecotherm}}
\label{fig:vapor-chamber}
\end{figure*}

For ISCR panels, the vapor chamber operates at approximately 25\textdegree C to 30\textdegree C, well within water's stable operating range at a vapor pressure of 0.032 to 0.042~bar. Key design parameters:

\begin{itemize}[nosep,leftmargin=1.5em]
    \item \textbf{Back face:} High-emissivity aluminum, 0.25~mm thick, forms the radiator surface
    \item \textbf{Top seal:} High-strength graphite-doped liquid crystal polymer (LCP), 0.3~mm
    \item \textbf{Pillars:} High thermal conductivity boron nitride spacers maintain 2~mm chamber thickness
    \item \textbf{Wick:} Copper or diamond/Cu mesh at $\sim$560~g/m\textsuperscript{2} provides capillary return
\end{itemize}

Vapor chamber technology has been demonstrated at heat flux densities exceeding 700~W/cm\textsuperscript{2} in the literature --- well above the $\sim$60~W/cm\textsuperscript{2} required here --- confirming substantial design margin. A rough thermal analysis has shown about a 5~degree delta from the 200\,mm~$\times$~70\,mm compute module dissipating 1~kW to our 1.7\,m square radiator.

Our vapor chamber must be designed to tolerate a transition from one bar external pressure before launch to external vacuum over a period of minutes. A small pressure relief valve in each panel could allow an initial fill of helium gas to escape until pressure is at the desired level of $\sim$0.04~bar. Then sufficient water is released from a liquid reservoir to provide proper function.

\subsection{Compute Performance of GPUs vs.\ Temperature}

To quantify the effect of temperature on cutting edge GPU performance, we extrapolated the 3\,nm fabricated NVIDIA Rubin GPU (1800\,W TDP) performance from its published operating conditions (45\textdegree C liquid cooling, estimated 85--90\textdegree C junction temperature, 0.82\,V core supply) to values of interest for supply voltage and temperature. Alternate lower power compute and module ICs could be expected to have similar behavior with junction temperature.

Table~\ref{tab:gpu} highlights how the voltage floor is lowered in cooler optimizations to reduce switching energy while increasing clock rates somewhat. Higher clock rates reduce the number of GPUs needed. Lower energy/token reduces required solar cell and radiator performance. TSMC provides different V\textsubscript{th} options for the N3 node, allowing ICs to be binned into these specific thermal-voltage profiles. Lower V\textsubscript{th} options can be used along with lower supply voltage at reduced temperatures. Four of these V\textsubscript{th} options are listed here. Use of higher V\textsubscript{th} transistors is required as temperature increases to keep static leakage current under control. Power from static leakage is converted to energy per token to sum with dynamic energy for the total energy per token calculation in Table~\ref{tab:gpu}. Two V\textsubscript{th} versions are listed for 25\textdegree C to demonstrate the balance between static and dynamic energy possible at the same temperature.

\begin{table*}[!ht]
\centering
\caption{Integrated performance and energy/token. \textit{Note: Table~4 values are generated by extrapolating the 3\,nm Rubin architecture away from the 45\textdegree C coolant temperature published data point. Non-45\textdegree C values would require experimental validation.}}
\label{tab:gpu}
\small
\begin{tabular}{cclcccccc}
\toprule
\makecell{\textbf{Coolant/}\\\textbf{Radiator}\\\textbf{Temp}} & \makecell{\textbf{Junction}\\\textbf{Temp (T\textsubscript{j})}} & \textbf{Cooling} & \makecell{\textbf{Transistor}\\\textbf{Flavor}\\\textbf{(N3P)}} & \makecell{\textbf{V\textsubscript{dd}}} & \makecell{\textbf{Clock}\\\textbf{Speed}} & \makecell{\textbf{Dynamic}\\\textbf{Energy}\\\textbf{(CV\textsuperscript{2})}} & \makecell{\textbf{Static}\\\textbf{Leakage}} & \makecell{\textbf{Total}\\\textbf{Energy/}\\\textbf{Token}} \\
\textdegree C & \textdegree C & & V\textsubscript{th} & volts & GHz & joules & joules & joules \\
\midrule
25 (ULVT) & $\sim$30--32 & vapor chamber & ULVT & 0.64 & 2.78 & 0.138 & 0.032 & 0.170 \\
25 (LVT) & $\sim$30--32 & vapor chamber & LVT & 0.69 & 2.72 & 0.149 & 0.022 & 0.171 \\
35 & $\sim$40--42 & vapor chamber & LVT & 0.74 & 2.6 & 0.170 & 0.034 & 0.204 \\
45 (baseline) & $\sim$85--90 & liquid & \makecell{SVT (Rubin\\nominal)} & 0.82 & 2.38 & 0.195 & 0.018 & 0.213 \\
\makecell{60 (max\\non-throttled)} & $\sim$103--105 & liquid & HVT & 0.85 & 2.05 & 0.209 & 0.065 & 0.274 \\
85 & \makecell{105\\throttled} & liquid & HVT & 0.88 & 1.35 & 0.224 & 0.098 & 0.322 \\
\bottomrule
\end{tabular}
\end{table*}

Key observations:

\begin{itemize}[nosep,leftmargin=1.5em]
    \item \textbf{Junction delta T to coolant:} can be much smaller using an efficient vapor chamber.
    \item \textbf{ISCR advantage:} By using an efficient vapor chamber (only available in a distributed architecture), IC junction temperature near 40\textdegree C allows targeting the LVT bin with improved performance over nominal 85--90\textdegree C junction T. This includes derating coolant T to 35\textdegree C from the ideal back-face temperature of $\sim$22\textdegree C (at 1000\,km SSO).
    \item \textbf{60\textdegree C is the maximum estimated coolant T allowed:} without reducing clock rate to maintain junction T within the Nvidia required 105\textdegree C (throttling).
    \item \textbf{Clock rate:} can be $>$25\% higher at 35\textdegree C vapor chamber T compared to 60\textdegree C liquid T.
    \item \textbf{The higher T energy penalty:} Staying stable near 105\textdegree C junction T requires high V\textsubscript{th} transistors (needing a supply voltage of 0.88\,V) to suppress leakage-induced instability, creating a double penalty: higher dynamic energy per switch (CV\textsuperscript{2}) and higher static leakage.
    \item \textbf{Even in the 60\textdegree C coolant optimized model:} leakage adds about 30\% of the dynamic energy value to the per token total. Total energy per token increases $>$30\% from 35\textdegree C to 60\textdegree C coolant temperature.
\end{itemize}

\section{Detailed Description of the ISCR Panel}

Given the background above, we describe a representative ISCR panel with 1~kW of power available for computation and communication to adjacent panels (wired) and the hub (fiber optic).

\begin{figure}[!ht]
\centering
\figframe{\includegraphics[width=0.95\columnwidth]{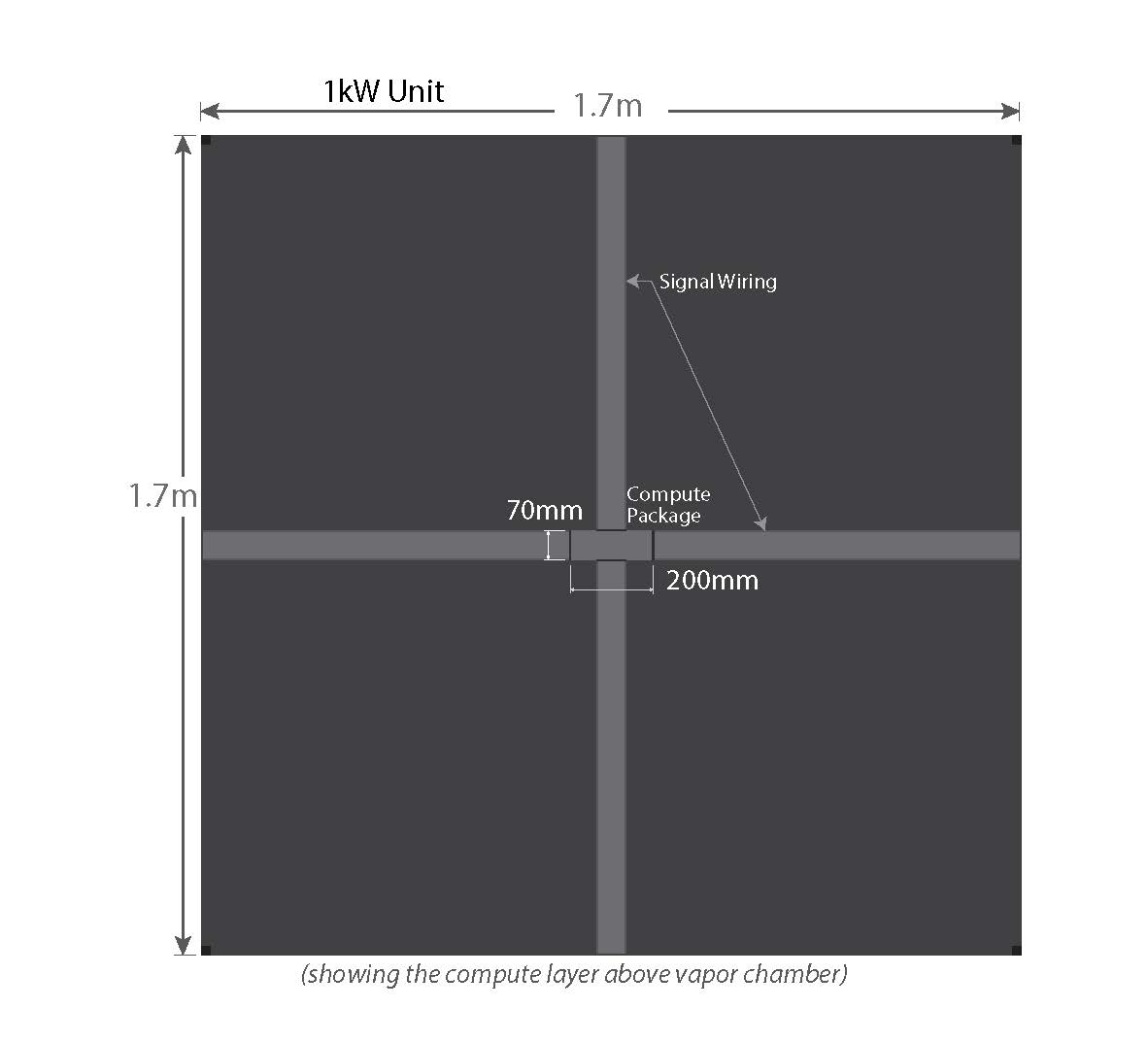}}
\caption{Integrated Solar Compute Radiator Panel --- top view showing the 1\,kW 1.7\,m~$\times$~1.7\,m panel. The compute package (70\,mm~$\times$~200\,mm) sits at the central intersection.}
\label{fig:panel-top}
\end{figure}

\subsection{Solar cell performance}

The solar cells are insulated from the backside as much as feasible while maintaining acceptable temperature for the cells. Depending on cell efficiency, this may be near total insulation via a vacuum gap. The values shown in Figure~\ref{fig:solar-efficiency} assume the solar cells are efficient enough to permit full insulation from the backside to achieve the lowest backside temperature possible. That means the waste heat from the solar cells must radiate from the front side.

The iROSA triple junction GaAs cells demonstrated recently on the ISS appear to have the needed performance, even though efficiency is expected to degrade from 31\% to 27\% at an 87\textdegree C frontside temperature (Table~\ref{tab:temps}). But the cost of GaAs cells may be prohibitive, so our design will focus on thin Si/Perovskite tandem cells with performance shown in Table~\ref{tab:solar}. The needed lower operating temperature ($\sim$70\textdegree C) is achieved by allowing some waste heat to leak to the back side, raising its temperature to about 35\textdegree C. The panel area required to provide 1~kW to the compute functions under these conditions is 2.9~m\textsuperscript{2}.

\subsection{Mechanical structure}

Solar cells are supported by the radiator structure, so can be very thin and light. The support is through sparse thermally insulating pillars. Vacuum provides insulation elsewhere with a total heat leak budget of 80\,W/m\textsuperscript{2} over a delta T of $\sim$30 degrees. The output current of several series cells goes to the compute module at voltages in the range of 10 to 50\,V to avoid thermal leakage from higher current conductors.

\begin{figure*}[!ht]
\centering
\figframe{\includegraphics[width=0.85\textwidth]{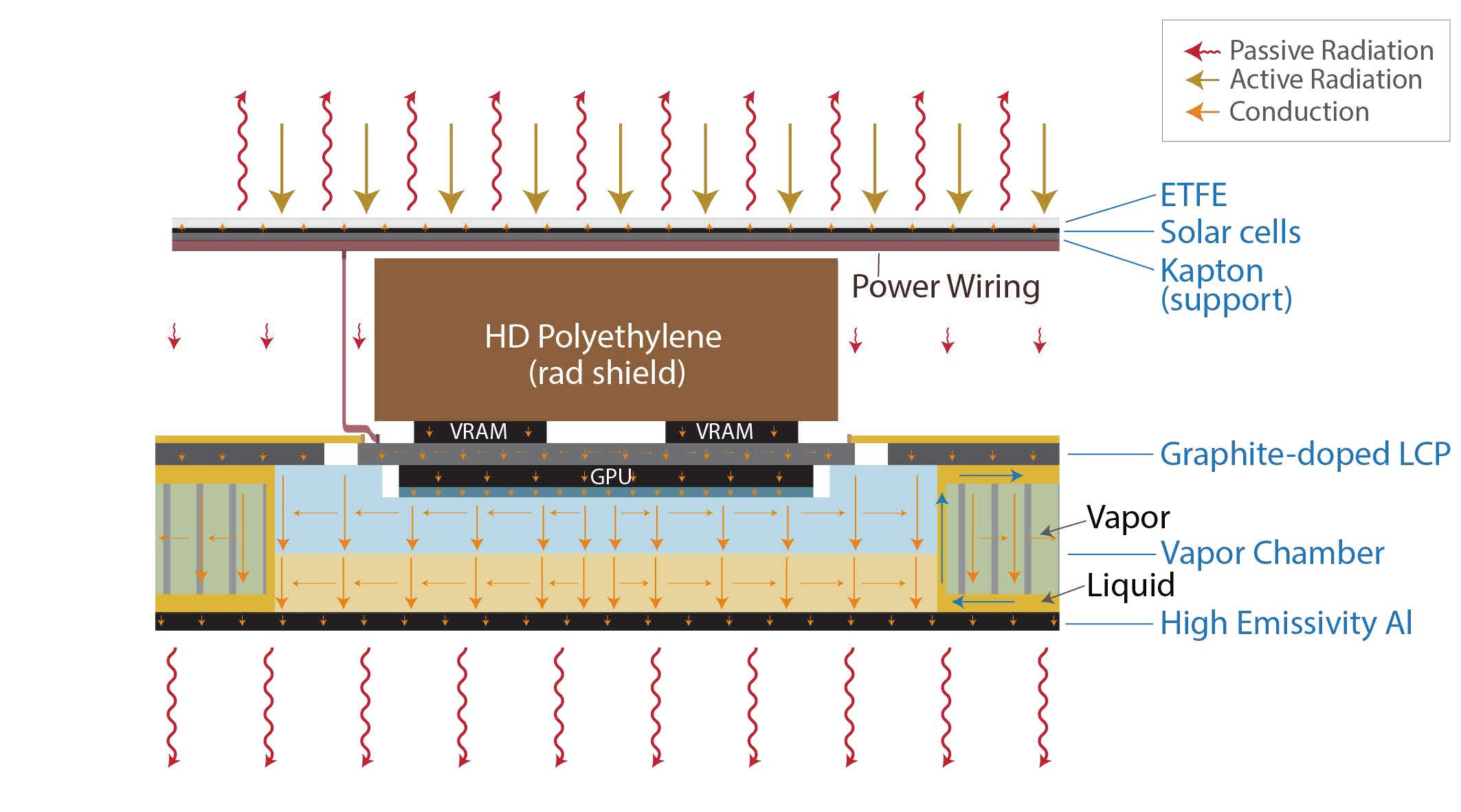}}
\caption{Integrated Solar Compute Radiator Panel --- cross-section view with the thickness exaggerated to show layer details.}
\label{fig:panel-cross}
\end{figure*}

We limit the compute module size in the rolled dimension to 70\,mm to avoid it interfering with the 2.5\,m roll radius of curvature. Limiting thickness to $<$7\,mm allows good stowed density. The width dimension is only constrained by mass considerations. We expect a 200\,mm width to be sufficient to contain all required circuitry and provide sufficient area for 1~kW of heat transfer to the vapor chamber.

Thinned silicon custom AI inference and memory ICs keep thickness low and help with heat transfer. They have high density interconnects through a silicon interposer. A 3\,mm layer of polyethylene is added to the top for proton shielding.

\subsection{Heat transfer}

This thin assembly mounts to a SiC heat spreader and ionizing radiation bottom shield that provides a large surface area (200\,mm~$\times$~70\,mm) interface to the vapor chamber. The vapor chamber operates at about 25\textdegree C allowing the use of water with its high thermal capacity at a vapor pressure of 0.04~bar that avoids the need for extra mass for structural strength. The backside of the radiator is high emissivity aluminum 0.25\,mm thick. The top is high strength graphite-doped liquid crystal polymer (LCP) sealed to the aluminum at the edges. High thermal conductivity boron nitride spacers maintain the 2\,mm thickness of the chamber. Wicking material inside the chamber masses 560~g/m\textsuperscript{2} if it is the commonly used copper mesh. The chamber efficiently transports heat distances up to about one meter, allowing panels as large as 2\,m~$\times$~2\,m.

\begin{figure}[!ht]
\centering
\figframe{\includegraphics[width=0.95\columnwidth]{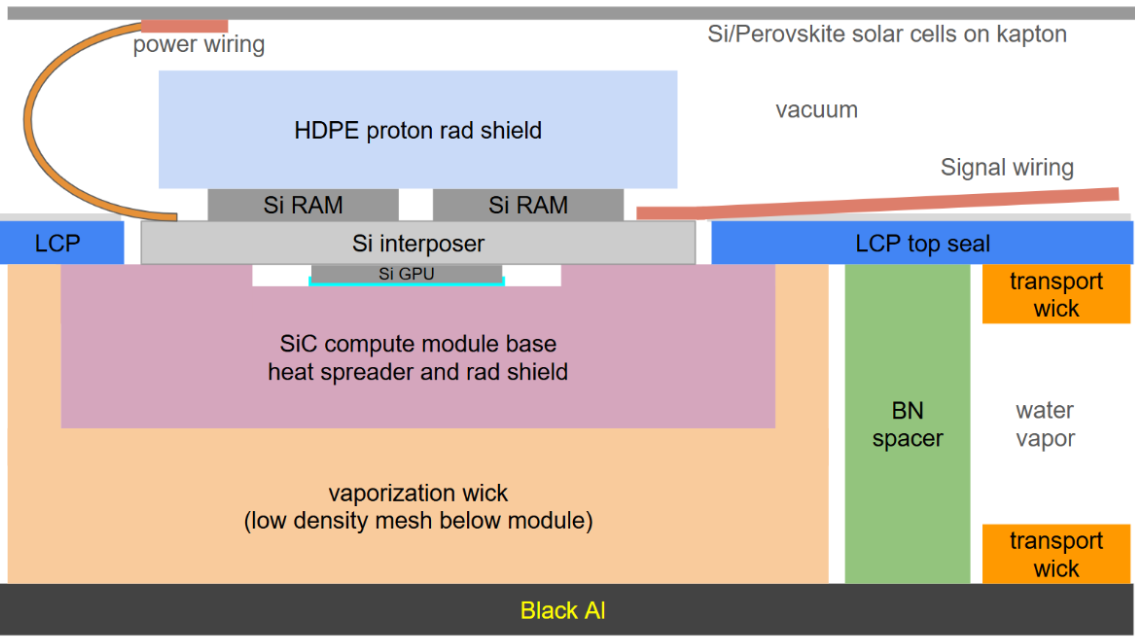}}
\caption{Conceptual cross-section of a panel with thin solar cells, the compute module (shielded from ionizing radiation), and the vapor chamber interface. Total thickness shown (not precisely to scale) is $<$7\,mm.}
\label{fig:compute-cross}
\end{figure}

The panel cross-section from front to back is: solar cells (0.2~mm) $\rightarrow$ thermal gap (4~mm, with sparse insulating pillar supports) $\rightarrow$ thermal reflector (Al, 0.01~mm) on graphite-doped LCP top (0.3~mm) $\rightarrow$ vapor chamber (2~mm) $\rightarrow$ high emissivity Al radiator (0.25~mm). Boron nitride spacers maintain vapor chamber thickness and provide front to back thermal conductivity. Total panel thickness: 6.4~mm. Estimated delta T from IC junctions: 5 to 7 degrees.

\subsection{Data communication capacity and power}

ISCR panels are organized in an array, described in Section~5. A 100~GB/s duplex data link to each adjacent panel is assumed in the wiring mass estimate using thin silver plated copper traces on thin polymer. There is abundant space available to widely separate lanes to avoid crosstalk. Eight lanes of NVLink or UA Link (802.3dj PHY) are assumed at the relatively low speed of 12~GB/s each to keep data error risk low for up to 2\,m distance.

This wiring is very low mass, so a higher data volume could be accommodated by adding lanes. Communication power per panel is estimated (assuming highly optimized custom ICs) at about 10\,W, including both the 100~GB/s wired links and one fiber optic link to the central hub.

\subsection{Mass density}

Total mass density is estimated at 3.1~kg/m\textsuperscript{2}, as shown in Table~\ref{tab:mass}. Note that while radiator mass area density dominates the total density, all of the panel structure and roll-out mechanism is included in the radiator mass budget. The total area mass density including radiator, compute, and solar cells is projected to be less than the iROSA density for just solar cells.

A 1\,m-diameter inflatable pneumatic stiffener runs along the panel edge only. Panels are designed to tolerate 6 to 10\,g acceleration in one direction to facilitate launch by Starship or a gentle lunar mass driver. 30\,g or higher may be feasible without significantly increasing mass.

\begin{table}[!ht]
\centering
\caption{Panel mass and thickness budget by material.}
\label{tab:mass}
\scriptsize
\begin{adjustbox}{max width=\columnwidth}
\begin{tabular}{llllccc}
\toprule
\textbf{Material} & \textbf{Dens.} & \textbf{Use} & \textbf{Area} & \textbf{Vol.} & \textbf{Mass} & \textbf{Th.} \\
 & (g/cm\textsuperscript{3}) & & (cm\textsuperscript{2}) & (cm\textsuperscript{3}/m\textsuperscript{2}) & (kg/m\textsuperscript{2}) & (mm) \\
\midrule
Si & 2.33 & solar cells & 9500 & 57 & 0.13 & 0.06 \\
ETFE & 1.75 & cell cover & 9500 & 133 & 0.23 & 0.14 \\
Kapton & 0.95 & support & 9500 & 95 & 0.09 & 0.1 \\
Al & 2.7 & power wiring & 950 & 23.8 & 0.06 & 0.25 \\
\textbf{Solar total} & & & & & \textbf{0.52} & \\
\midrule
HD polyethylene & 0.95 & proton shield & 130 & 39 & 0.04 & 3 \\
Si & 2.33 & ICs, interposers & 65 & 3.9 & 0.01 & 0.6 \\
SiC & 2.3 & PCB, heat spread & 130 & 39 & 0.09 & 3 \\
Cu & 8.96 & inter-panel wiring & 3000 & 8 & 0.07 & 0.008 \\
Ag & 10.49 & plating for HF & 3000 & 2 & 0.02 & 0.002 \\
\textbf{Compute total} & & & & & \textbf{0.23} & \\
\midrule
Graph.-doped LCP & 1.5 & vapor ch.\ top & --- & 300 & 0.45 & 0.3 \\
Boron nitride & 2.2 & vapor ch.\ pillars & 800 & 160 & 0.35 & 2 \\
Aluminum & 2.7 & radiator, vp ch.\ bot & 10000 & 250 & 0.675 & 0.25 \\
Diamond/Cu mesh & 2.5 & wick & 9000 & 225 & 0.56 & 0.25 \\
Water & 1 & working fluid & 9000 & 180 & 0.18 & 0.2 \\
\makecell[l]{Carbon-fiber reinf.\\polycarbonate} & 1.3 & flexible structure & --- & 100 & 0.13 & --- \\
Argon (STP) & 0.0018 & pneu.\ stiffener & --- & 10000 & 0.02 & 1000 \\
\textbf{Radiator total} & & & & & \textbf{2.4} & \\
\midrule
\textbf{Total} & & & & & \textbf{3.1} & \textbf{6.4} \\
\bottomrule
\end{tabular}
\end{adjustbox}
\end{table}

\section{Implementation of Inference LLM}

This section briefly reviews LLM design and details several LLM implementations that might run on the ISCR array.

\subsection{Review of LLM Parallel Distribution}

We choose and analyze an inference-only LLM as a representative best use of the ISCR array, because it has lighter inter-GPU requirements than a training implementation. We assume an array of the already-described ISCR panels, each with 1\,kW power supporting a GPU of 1,000~TFLOPs capacity, reflecting a near term expectation of 1~TFLOP/W for the best terrestrial GPUs. Full ISCR array description follows in Section~6.

A large language model (LLM) consists of a series of \textbf{attention blocks}, each performing heavy matrix computations on the current token sequence. The model also maintains a \textbf{context} --- the running token history --- of 100,000 to 1,000,000 tokens comprising the user/LLM interaction thus far. Each new token's computation takes into account that token history, so longer contexts dramatically increase both compute load and memory requirements. An LLM also has up to 1 trillion numerical weights totalling perhaps 2~TBytes, and a per-attention-block history of prior per-token computations called a \textbf{KV cache}, may be up to 4~MB per attention block, best held in small, fast memory.

In practice, computation of a single token of output may thus require terabytes of memory, and trillions of computations. So, distribution of the work across multiple GPUs is essential. This is done two ways for inference LLM operations.

\textbf{Tensor parallelism} distributes the matrix work of a single attention block across multiple GPUs. The GPUs can share the expensive attention computations and the KV cache. But, tensor parallelism requires high inter-GPU bandwidth and thus is typically confined to within a single server or rack in terrestrial deployments.

\textbf{Pipeline parallelism} instead divides the sequence of attention blocks into stages, with each GPU group handling one or more blocks and passing activations to the next stage. This requires far less bandwidth between stages --- making it well-suited to the wired inter-panel links of the ISCR array.

Both parallelism strategies require multiple LLM sessions \textbf{in flight simultaneously}, so that all pipeline stages always have active work. Memory constraints are significant: the up to one trillion weights of a large LLM must be distributed across GPUs, but a deep pipeline requires more concurrent in-flight sessions, increasing KV-cache memory demand per GPU.

\subsection{Organization of Sample Inference LLMs}

ISCR panels can be organized in \textbf{4-panel squares} (2$\times$2 arrangements), each forming the functional equivalent of a terrestrial rack-mounted server. Within each quad, four GPU panels connect via 100~GB/s (including link overhead) wired links, enabling 4-way tensor parallelism with low latency and $\sim$1.65~GB/s tensor bandwidth --- comparable to NVLink within a server rack. Quads are then chained into a \textbf{pipeline}, with each quad or single panel handling one or more attention blocks, and passing activations to the next via the same inter-panel links.

Table~\ref{tab:llm} summarizes four sample LLM inference implementations. The columns shown are: Context Length, Pipeline Length, P-parallel (pipeline stages), T-parallel (tensor-parallel GPUs per stage), All memory per GPU (GB of HBM), and All communication bandwidth per GPU (duplex GB/s).

\begin{table}[!ht]
\centering
\caption{Sample LLM Inference Implementations on the ISCR Array.}
\label{tab:llm}
\footnotesize
\begin{adjustbox}{max width=\columnwidth}
\begin{tabular}{lcccccc}
\toprule
\textbf{Model} & \makecell{\textbf{Context}\\\textbf{Length}} & \makecell{\textbf{Pipe.}\\\textbf{Length}} & \makecell{\textbf{P-}\\\textbf{par.}} & \makecell{\textbf{T-}\\\textbf{par.}} & \makecell{\textbf{All Mem/}\\\textbf{GPU (GB)}} & \makecell{\textbf{All BW/}\\\textbf{GPU (GB/s)}} \\
\midrule
\makecell[l]{16-panel, no\\tensor parallelism} & 100,000 & 96 & 16 & 1 & 46.45 & 0.55 \\
\addlinespace
\makecell[l]{16-panel w/tensor\\parallelism} & 100,000 & 96 & 4 & 4 & 46.45 & 2.2 \\
\addlinespace
384-panel light LLM & 100,000 & 96 & 96 & 4 & 1.94 & 52.84 \\
\addlinespace
512-panel heavy LLM & 500,000 & 128 & 128 & 4 & 7.06 & 14.51 \\
\bottomrule
\end{tabular}
\end{adjustbox}
\end{table}

\begin{itemize}[nosep,leftmargin=1.5em]
    \item \textbf{``16-panel, no tensor parallelism''} offers the smallest reasonable footprint, at 672 tokens/sec per in-flight inference across 32 simultaneous sessions, with no tensor communication overhead. A viable minimum viable configuration.
    \item \textbf{``16-panel w/tensor parallelism''} delivers the same small footprint but quadruple throughput (2,688 tokens/sec) on only 8 in-flight sessions --- the 4-way tensor parallelism within each panel quad provides server-like performance with inter-panel copper links.
    \item \textbf{``384-panel light LLM''} is a maximum-footprint implementation requiring much less per-GPU memory (1.94~GB vs.\ 46.45~GB), but significantly higher communication bandwidth (52.84~GB/s duplex) --- feasible given the 100~GB/s per-panel wired links.
    \item \textbf{``512-panel heavy LLM''} implements a 500,000-token context with 128 pipeline stages. It achieves 553 tokens/sec across each of 256 simultaneous in-flight inferences, with moderate memory (7.06~GB/GPU) and moderate bandwidth (14.51~GB/s). This is the configuration described in the Abstract: 16 such subarrays on a full satellite yield $>$4,000 concurrent inferences.
\end{itemize}

Compute efficiency (discussed in section 5.2) does not seem to suffer significantly by limiting power/panel to 1\,kW, and a highly optimized inter-panel communication may need $<$10\,W. Limiting panel area to 2.9\,m\textsuperscript{2} is helpful for vapor chamber efficiency.

\section{Design of an ODC Satellite using the ISCR panel}

\subsection{Deployment}

The ISCR array is designed to facilitate deployment by roll out in a manner somewhat similar to ROSA solar arrays, but with a minimum roll radius of curvature of 2.5\,m. The complete satellite makes efficient use of a full single Starship V3/V4 payload volume (8\,m diameter cylinder 22\,m long, extended length version).

In the primary design, panels are wound around a 5\,m-diameter central bus cylinder to an outer coil radius of $<$8\,m, fitting within the 9\,m Starship diameter. The 5\,m diameter hub provides room for relatively low mass pressure vessels containing argon (propellant and pneumatic array deployment gas).

Deployment is pneumatic: pressurized argon inflates braided Kevlar/polyimide tubes along each panel rim and transverse spines, unrolling the array without motors, springs, or pyrotechnics. The tubes will likely contain multiple cells so that differential pressure can be adjusted to insure a flat array.

Communication, command and control functions, and most of the ion propulsion system also reside in the hub. Some small ion thrusters would be placed at the array ends for rotation control.

\begin{figure*}[!ht]
\centering
\figframe{\includegraphics[width=0.9\textwidth]{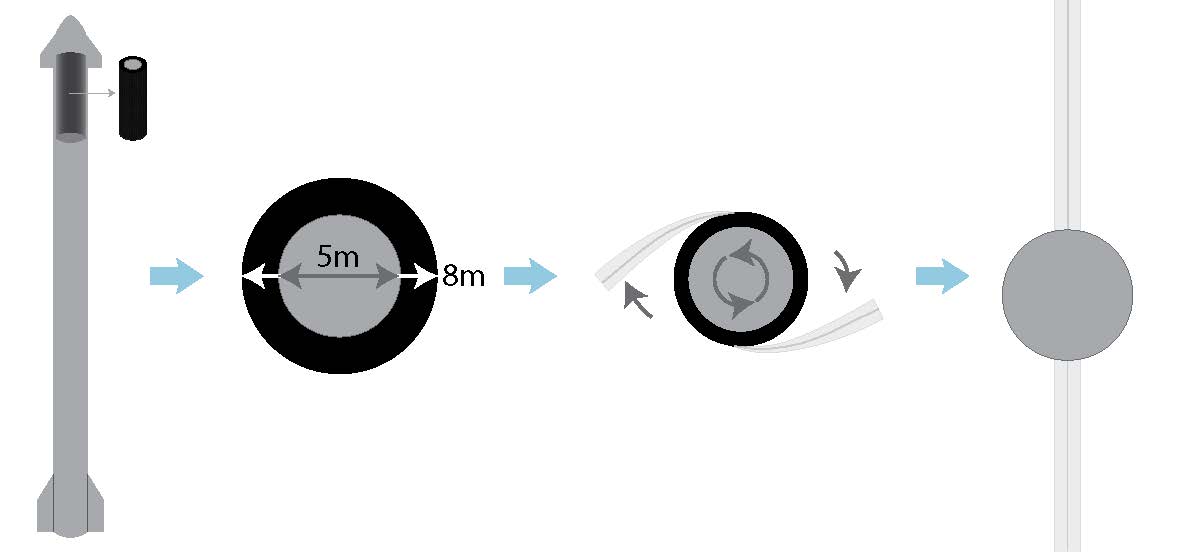}}
\caption{Diagrams of the ISCR array --- single-bus stowage (panels wound around 2.5\,m-radius central bus, outer radius up to 4\,m), deployment sequence, and deployed configuration (22\,m~$\times$~2200\,m). The dual-bus alternative places two half-cylinder buses at the array ends.}
\label{fig:deployment}
\end{figure*}

At our design panel thickness (10~mm including roll layer spacers), the volume density is sufficient for $>$60,000~m\textsuperscript{2} of solar array area within the payload volume and could provide over 22~MW of compute power. However, that would require $\sim$186 tons of array mass. If the mass of the hub is $<$14 tons, that would fit within the SpaceX payload goal of 200 tons.

\subsection{Mass and specific power}

To allow mass margin we are constraining the design to 150 tons. At that mass, $\sim$16~MW would be available for compute functions, allowing $\sim$4\% for other functions. Each of 16,600 panels of 1~kW output are 2.9~m\textsuperscript{2}, 1.7\,m on a side if square. Rows of 12 panels fit within 20.4\,m of cylinder length. The total length of the rolled out arrays would be $\sim$2.2~km, and panel area = 45,000\,m\textsuperscript{2}.

\begin{table}[!ht]
\centering
\caption{ISCR satellite mass with stow limit and specific power.}
\label{tab:satmass}
\small
\begin{tabular}{lr}
\toprule
\textbf{Total array area density, kg/m\textsuperscript{2}} & \\
\midrule
solar array & 0.52 \\
compute & 0.23 \\
radiators & 2.40 \\
total area density & \textbf{3.15} \\
mass if limited by stow density, tons & \textbf{197} \\
\midrule
\textbf{Total distributed mass, tons} & \textbf{141.8} \\
communication and control mass & 5\% \\
\textbf{Total satellite mass, tons} & \textbf{148.8} \\
communication and control power & 4\% \\
\textbf{Specific compute power kW/ton} & \textbf{112.5} \\
\bottomrule
\end{tabular}
\end{table}

\begin{figure}[!ht]
\centering
\figframe{\includegraphics[width=0.95\columnwidth]{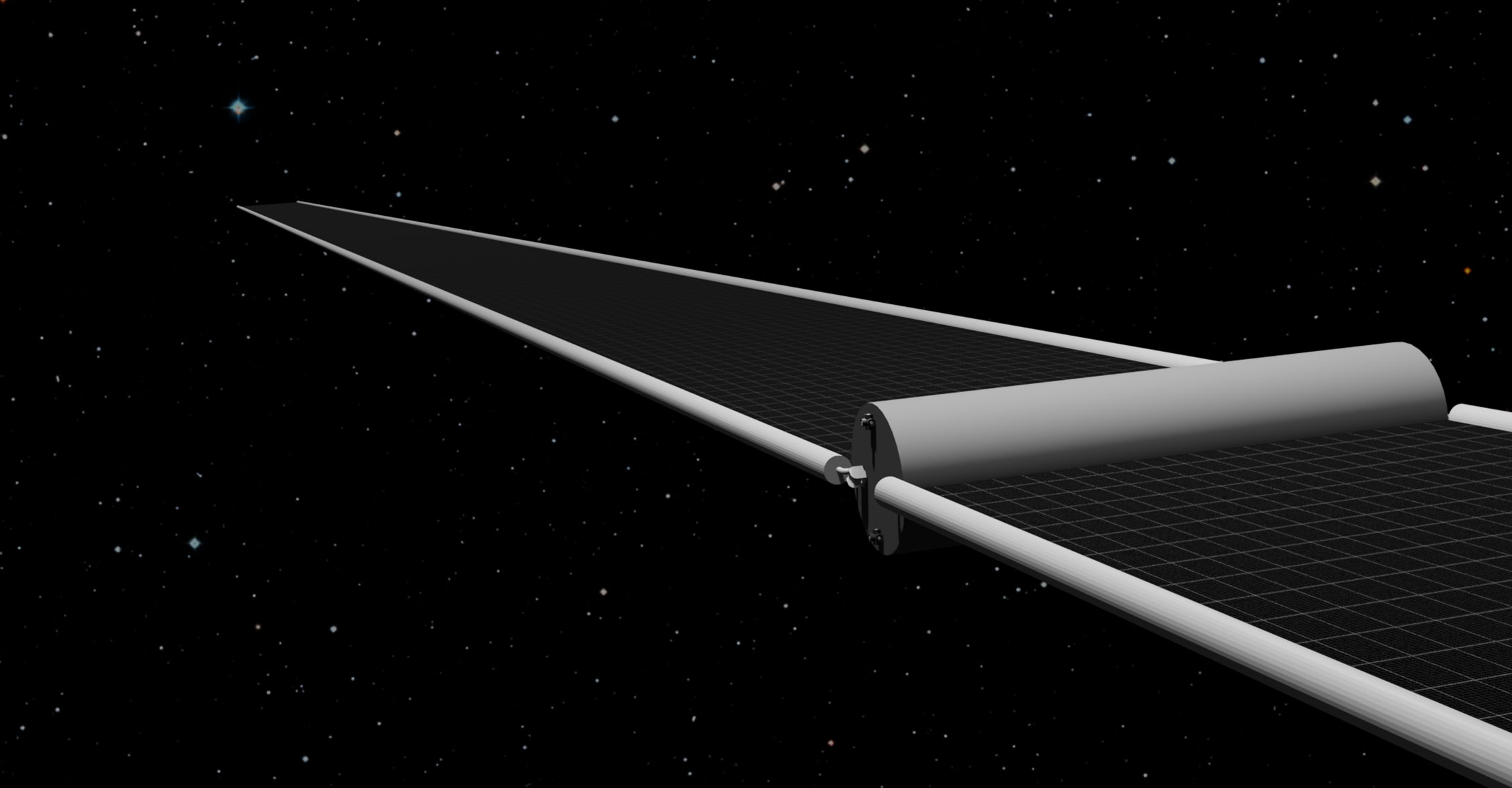}}
\caption{Deployed satellite with pneumatic stiffeners at edges.}
\label{fig:deployed}
\end{figure}

The central bus ($\sim$5.5~tons) houses attitude control hardware, two Busek BHT-600 Hall thrusters (39~mN each) for drag compensation at LEO altitudes, an argon inflation system, and command/data handling.

\begin{figure*}[!ht]
\centering
\figframe{\includegraphics[width=0.9\textwidth]{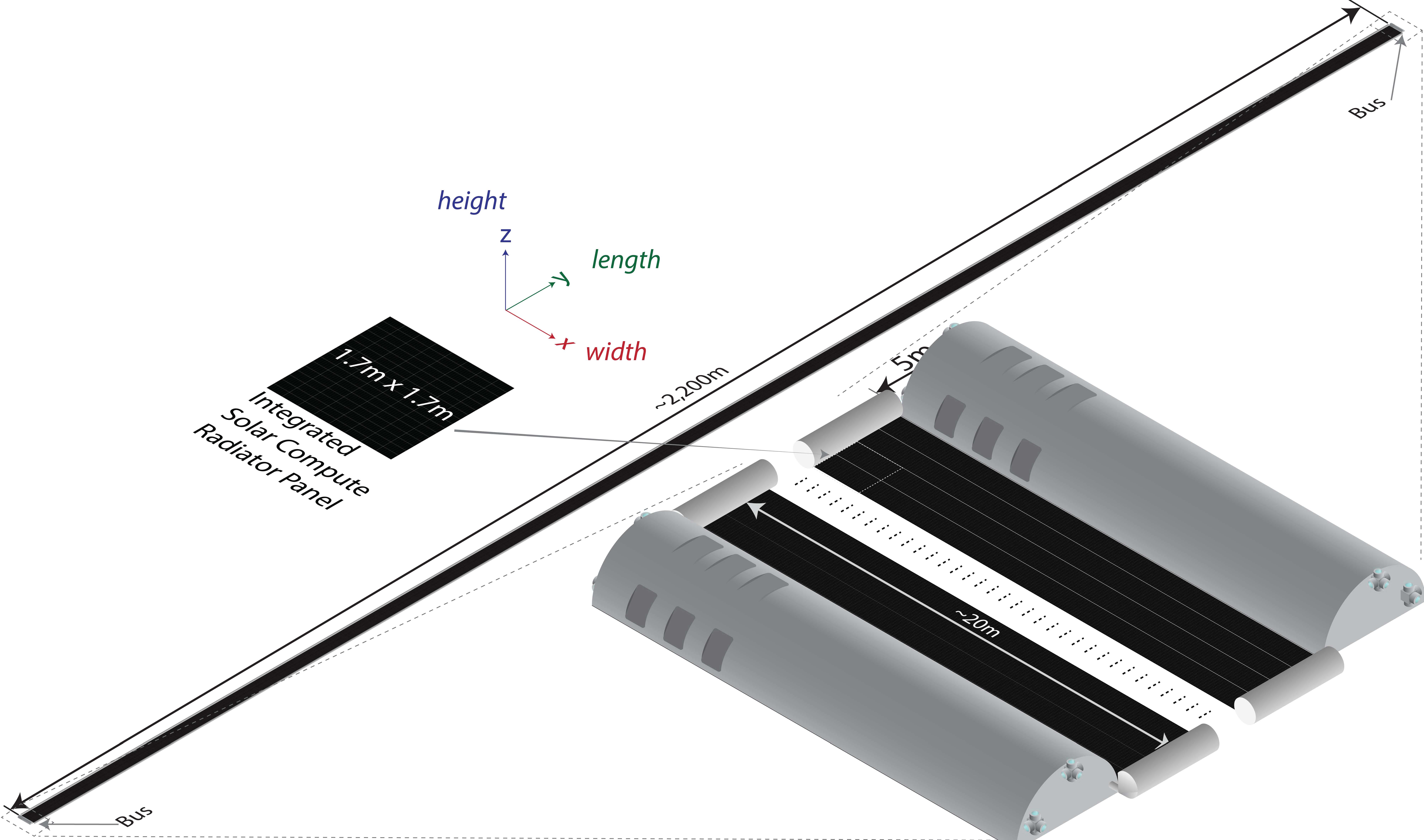}}
\caption{An alternative deployment design places two half-cylinder buses at opposite ends of the deployed array, improving structural stiffness and allowing independent thrust on each end of the array.}
\label{fig:dual-bus}
\end{figure*}

\subsection{Scalability at, and beyond, maximum Starship payload}

The distributed architecture scales more favorably to larger satellites than conventional monolithic designs:

\begin{itemize}[nosep,leftmargin=1.5em]
    \item Distributed power generation, consumption, and radiation avoids the need to transport MWatts of electrical power over km distance and thermal power over hundreds of meters distance. Data signals can easily be transported by fiber optics.
    \item At maximum Starship payload scale, a single large satellite carries less control and communication overhead per unit power than a dozen smaller satellites each requiring independent bus hardware.
    \item At constellation scale, reaching 1~TW of compute would require 1,000,000~$\times$~1~MW satellites with conventional designs, but only 100,000~$\times$~10~MW satellites with the distributed Integrated Panel approach --- a 10$\times$ reduction in bus overhead.
    \item Panel failure at scale is handled gracefully: a failed panel reduces available pipeline stages or tensor-parallel capacity but does not halt computation. Pipeline reconfiguration routes around failures, and the large panel count (tens of thousands) provides strong statistical resilience.
    \item Increasing satellite size reduces per-unit launch overhead, improving economics at scale.
    \item Less concern for end of life and unplanned demise of an ISCR satellite with its mostly low mass/area density.
\end{itemize}

\begin{figure}[!ht]
\centering
\figframe{\includegraphics[width=0.95\columnwidth]{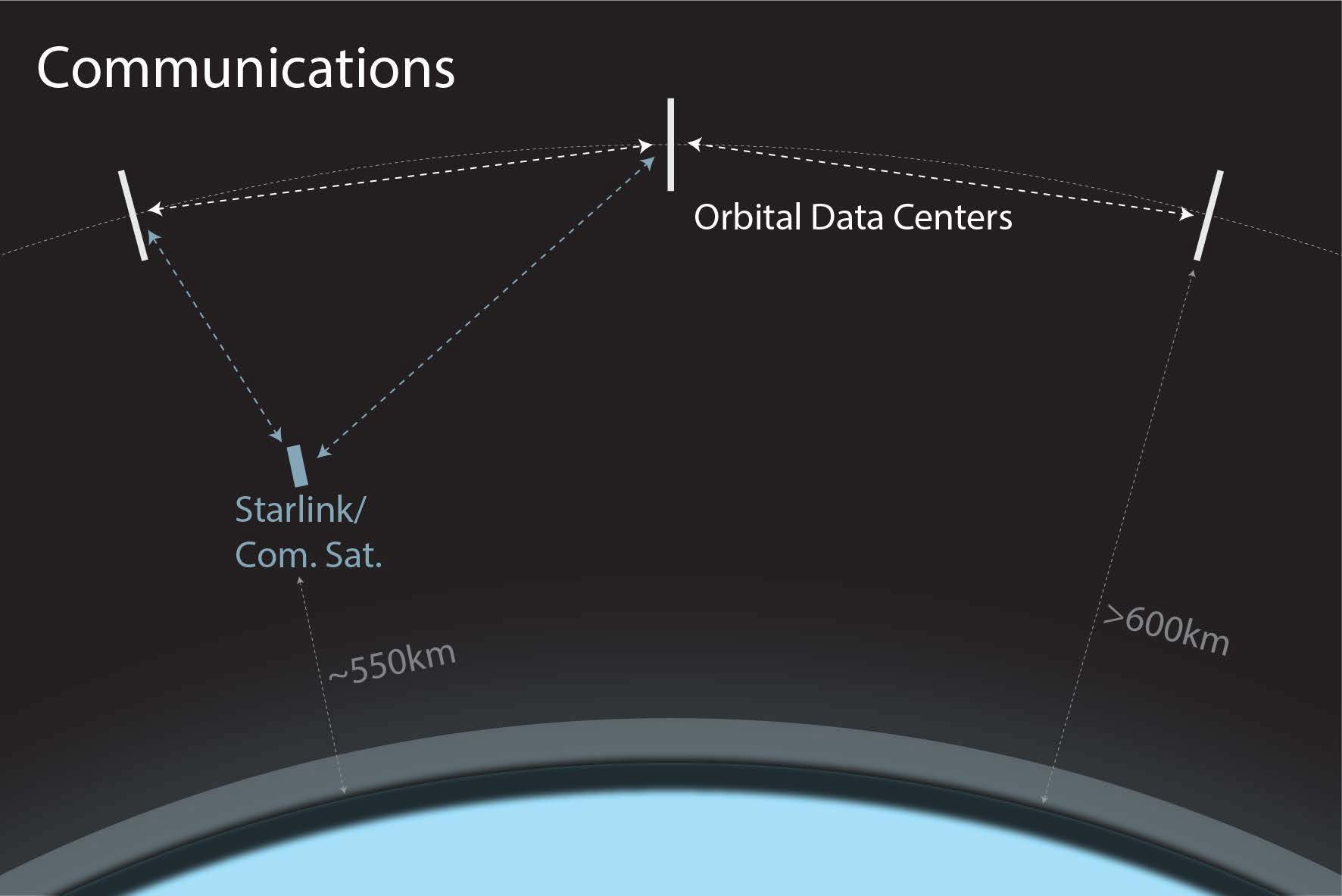}}
\caption{ISCR ODC Communications architecture --- inter-satellite links between orbital data centers at $>$600\,km, with relay through Starlink/communication satellites at $\sim$550\,km.}
\label{fig:comms}
\end{figure}

ISCR panels are mass-producible on continuous manufacturing lines, analogous to flexible solar module and PCB production. This continuous-line manufacturability is a key cost driver: at the required scale of $\sim$16,000 panels per satellite, batch manufacturing economics apply.

\section{System Performance Advantages vs.\ Conventional ODC}

This section compares the ISCR integrated energy panel with distributed compute architecture against conventional separate-subsystem ODC satellite designs (not terrestrial installations) including low and high temperature radiator versions.

The ultimate comparison should be in cost per compute metric. But that is difficult given the cost uncertainty for major elements such as solar cells, radiators, compute ICs, compute packaging, and finally launch cost. So until these parameters are better defined, some cost/impact trades are presented to help with system comparisons.

\subsection{Specific power}

We estimate that despite its large radiator, the ISCR system can provide specific power of 112~kW/ton (Table~\ref{tab:satmass}), which is within the SpaceX goal of 100 to 150~kW/ton.

The primary structural advantage of the ISCR architecture is the decision to use the vapor chamber radiator as the sole mechanical substrate for the solar cells. This eliminates a dedicated solar panel structure, yielding an array-level specific power of approximately 506~W/kg --- more than five times the $<$100~W/kg typical of existing satellites.

A common objection to distributed compute is the performance loss due to communication between panels (additional power and latency). We estimate that duplex 100~GB/s between panels is more than sufficient to avoid slowing computation. The power added by 400~GB/s (4 adjacent panels) is $<$10\,W per panel for a highly optimized design.

The distribution of computation across a wider area may negatively affect compute efficiency, but the design we offer closely mirrors terrestrial distribution with servers (tensor parallel) and interserver (pipeline parallel), and we estimate that sufficient bandwidth can be maintained inter-panel with proper design.

\subsection{Cost per computation}

Orbital compute satellites ultimately must be compared by cost per computation, not just power/mass. A step toward that is energy cost per computation (discussed as Joules/token in the background section). Here we attempt a quantitative summary of the potential performance of several designs mostly distinguished by cooling and radiator technology, and the effect each has on compute IC junction temperatures. We limit consideration of junction temperature T\textsubscript{j} to 105\textdegree C or less, as specified for most GPUs. Other temperatures are as described in section~3.3, except for the high performance radiator/cooling technology which we speculate may reduce the junction to radiator delta T to 25 degrees, probably at the cost of higher complexity and mass density.

Table~\ref{tab:perf8a} shows how a fraction of the absorbed power density is available for computation which must (along with IR absorbed from Earth) be radiated at the chosen temperatures of solar cells, IC junctions, and radiators. Operating at a junction temperature of 35\textdegree C provides some computation benefit over typical liquid cooled terrestrial systems operating at 45 degrees. That is enough for a small increase in clock rate over typical liquid cooling, and a large improvement over high radiator temperature designs. Lower temperatures also increase reliability, reducing post-deploy costs.

\begin{table*}[!ht]
\centering
\refstepcounter{table}\label{tab:perf8a}
\addtocounter{table}{-1}
{\small\textbf{Table 8a:} Performance comparison of ISCR, conventional architectures with low and medium radiator temperatures, and a speculated high performance radiator/cooling technology.}\par
\vspace{4pt}
\small
\begin{tabular}{clccccl}
\toprule
 & \textbf{System} & \textbf{ISCR} & \makecell{\textbf{Low T}\\\textbf{radiator}} & \makecell{\textbf{Medium T}\\\textbf{radiator}} & \makecell{\textbf{High T}\\\textbf{radiator}} & \\
\midrule
2 & Solar absorption & 82\% & 82\% & 82\% & 82\% & thin Si/Perovskite \\
3 & cell efficiency & 27\% & 30\% & 30\% & 30\% & \\
\addlinespace
 & \textit{Power to radiate, W/m\textsuperscript{2}} & & & & & \\
5 & compute power & 367 & 408 & 408 & 408 & \\
6 & Earth IR & 12 & 24 & 24 & 24 & \\
7 & solar cell transfer & 80 & & & & \\
\addlinespace
 & \textit{Temperatures, \textdegree C} & & & & & \\
9 & solar cells & 66 & 27 & 27 & 27 & \\
10 & \textbf{compute junctions} & \textbf{41} & \textbf{90} & \textbf{105} & \textbf{105} & \\
11 & \textbf{radiator} & \textbf{35} & \textbf{45} & \textbf{60} & \textbf{80} & \\
12 & radiated at T, W/m\textsuperscript{2} & 460 & 523 & 629 & 794 & \\
13 & radiator sides & 1 & 2 & 2 & 2 & \\
\bottomrule
\end{tabular}
\end{table*}

\begin{table*}[!ht]
\centering
\label{tab:perf8b}
{\small\textbf{Table 8b:} Figures of merit values associated with cooling technology choices.}\par
\vspace{4pt}
\small
\begin{tabular}{clcccccl}
\toprule
 & \textbf{Radiator} & \textbf{ISCR} & \makecell{\textbf{Low T}\\\textbf{radiator}} & \makecell{\textbf{Medium T}\\\textbf{radiator}} & \makecell{\textbf{High T}\\\textbf{radiator}} & & \textbf{Cost impacts} \\
\midrule
15 & \textbf{Radiator size (\% of solar array)} & \textbf{100\%} & \textbf{41\%} & \textbf{34\%} & \textbf{27\%} & & radiator mass \\
16 & cooling technology & vapor chamber & liquid & liquid & high performance & & radiator, mass \\
17 & \textbf{GPU clock rate, GHz} & \textbf{2.6} & \textbf{2.38} & \textbf{2.05} & \textbf{2.05} & & ICs \\
18 & energy per token & 0.204 & 0.213 & 0.274 & 0.274 & & solar cell, radiator \\
19 & \makecell[l]{\textbf{token normalized}\\\textbf{compute power, W/m\textsuperscript{2}}} & \textbf{384} & \textbf{408} & \textbf{317} & \textbf{317} & & solar cell, radiator \\
\bottomrule
\end{tabular}
\end{table*}

Table~8b shows how figures of merit vary with the cooling temperature choices and associated cooling technology. The rightmost column in Table~8b tracks which parameters impact various component mass and cost.

Row 15 of the table compares the required radiator size for each design choice (normalized to the solar array size). If launch costs were high and all radiators had the same mass density, the medium or high T designs would have considerable advantage. But the simplicity of the ISCR cooling could allow low enough mass density for it to be competitive on a radiator mass basis.

Row 17 shows that if the cost of GPUs was the totally dominant cost, the higher clock rates allowed by low junction temperatures would make the low compute temperature ISCR approach the clear choice.

Rows 18 and 19 show that if the cost of solar cells was the totally dominant cost, the token normalized (to Low T) compute power would favor the low T radiator, due to its solar cell efficiency and good compute efficiency, with the ISCR approach a close second. The energy per token advantage at low junction temperatures may increase as IC feature sizes decrease (see 7.3 below) making ISCR favored for solar cell cost as well.

Given that ``the best part is no part''~\cite{musk}, eliminating separate radiator and its moving parts could allow the passively cooled ISCR to be comparable or better in mass to minimized radiator designs, especially for large arrays. As launch costs decrease, the effect of any mass penalty of a larger radiator decreases. Also, as feasible launch mass increases, ISCR may offer much greater single-satellite payload scalability (including stowage density) than conventional designs with separate units for cooling.

\subsection{Thermal limits for Si ICs at 2\,nm and beyond}

As transistor densities have increased beyond the Dennard scaling era, power per mm\textsuperscript{2} has increased dramatically. 3\,nm and 2\,nm processors seem to be near a thermal limit. Going beyond 2\,nm and operating with increased supply voltage to accommodate high temperatures appears extremely difficult. Keeping junction temperatures below $\sim$80\textdegree C for scaling below 2\,nm seems necessary to benefit from increased computation per watt.

\subsection{Si processor and memory supply limitation}

The scale and rate of proposed ODC deployment may greatly exceed future production capacity of silicon foundries even with unprecedented capex. Just as electrical generation capacity is currently limiting terrestrial data center expansion, silicon wafer production may limit ODC growth. Under potential rationing conditions, the price of processors and memory ICs may significantly exceed the expected cost of production. Also, higher defect densities may become acceptable if they allow new capacity to be rushed into production.

If these concerns prevail, a distributed compute system that uses ICs that are small enough for high yield and clock at the highest rates would have an advantage.

\section{Future Work}

Several areas require further investigation for the ISCR architecture:

\subsection{Panel level investigations}

\textbf{Cooling system simulation.} Accurately simulate and optimize the thermal design of the compute module and the vapor chamber with 3D tools.

\textbf{Radiation qualification.} Compute ICs and Si/Perovskite thin-film cells require comprehensive qualification for $\sim$5 year LEO radiation environments. The planned 3\,mm of HDPE top shield of the compute module may be marginal. It can be increased to over 5\,mm without significantly increasing mass, but that could impact stowage density. Ground-based proton and electron irradiation experiments are needed.

Radiation shielding mass and volume trades against operational lifetime, which is also limited by obsolescence as higher performance compute ICs are developed. This design variable must be optimized with these factors in mind. Some values $<$5 years may be sufficient.

\textbf{Custom ASIC performance and NRE.} ISCR economics depend substantially on custom ASIC performance and amortization over high production volumes. Preliminary ASIC architecture definition --- memory hierarchy, inter-panel interface, power management --- is needed to bound NRE costs and identify the crossover production volume where not only custom silicon IC design, but custom silicon fabrication process, outperforms available merchant GPUs.

\textbf{2\,nm and sub-2\,nm fabrication.} The benefits of sub-3\,nm IC nodes for energy per token in the ISCR thermal environment have not been analyzed sufficiently. As TSMC N2P and beyond become available, the voltage-frequency-temperature optimization space should be revisited.

\textbf{Custom ASIC Availability and Performance.} Custom ASIC NRE costs are not included in the mass-specific figures above, but given the expected industrial scale, this is likely similar for most all approaches. Also, Table~3 (GPU temperature-performance) is extrapolated from a single operating point and requires experimental validation.

\subsection{Satellite level investigations}

\textbf{Structural statics and dynamics to maintain panel flatness.} The torsional structural statics and dynamics of the 1100\,m flexible panel array are uncharacterized. This is the highest-priority structural risk; a full analysis is needed and active pneumatic or thrust forces may be required. This investigation will also determine whether a single- or dual-bus design is best.

\textbf{Fault tolerance.} A large homogenous array of ISCR panels offers considerable opportunity for fault tolerance. Research into additional communication links, and an algorithm for factoring out faulty panels, would enable this.

\textbf{Orbital station keeping and transfer time.} At 39~mN thrust and $\sim$104 tons, LEO-to-1000\,km-SSO or LEO-to-L5 transfer times and propellant budgets require detailed trajectory analysis. Transfer time is a significant mission variable affecting both time-to-revenue and propellant mass allocation.

Tidal acceleration of 5~mm/s\textsuperscript{2} across a 2\,km structure should also help maintain radial orientation, possibly eliminating active station keeping for that dimension.

\textbf{Relative costs for solar cells, compute ICs, radiators, and launch mass:} As mentioned in Section~7.2, the relative importance of these costs will likely determine what architecture is most effective for orbital data centers. Further study of these costs is warranted.

\textbf{Comparison with prior work.} Competing designs may achieve comparable specific power regardless of subsystem optimization. Even if they preserve dedicated structural substrates for each subsystem, advances in solar cell substrates could be significant. This performance should be validated against future publications as the ODC field matures.

\section{Conclusion}

We have described and analyzed the Integrated Solar Compute Radiator (ISCR) architecture for orbital data centers featuring single panels offering equal radiation and solar cell areas with common support. The primary insights:

\begin{itemize}[nosep,leftmargin=1.5em]
    \item A 35\textdegree C compute junction temperature may offer $>$30\% improvement in energy per token over a high temperature radiator design (105\textdegree C junction T). This difference may increase at smaller IC fabrication feature sizes.
    \item A single large structure using the vapor chamber radiator as the mechanical substrate for the solar cells eliminates a dedicated solar panel structure entirely, allowing a solar cell array specific power of approximately 506~W/kg, five times existing conventional designs.
    \item System-level expected specific power is approximately 112~kW/ton including deploy and station keeping propulsion mass.
\end{itemize}

The resulting architecture enables a 150~ton single-launch satellite with 45,000~m\textsuperscript{2} of integrated solar/compute/radiator area, and 16.4~MW of solar power. The 1\,kW/panel design accommodates distributed LLM inference from a compact 16-panel configuration to a 512-panel array supporting 256 simultaneous in-flight inferences at 500,000 token context. The distribution of inference over several panels mirrors such distribution in terrestrial data centers and should not significantly degrade compute performance. Argon inflation deploys the full array without motors or pyrotechnics.

The economic case for the ISCR architecture benefits from improved compute performance at low temperature, and the low mass thin-film solar cell pathway the structural integration enables. Where the cost per compute metric trade between small high temperature radiator and low temperature compute ultimately falls depends on launch cost progress, thermal and compute simulations, radiation qualification results, and many other factors not yet available.

Priority future work: FEA analysis and design optimization of panel torsional statics and dynamics, 3D thermal simulation of the panel, monitoring thin Si/Perovskite LEO qualification including radiation testing, and ASIC architecture definition to understand costs vs design and fabrication node.


\end{document}